\documentclass[12pt]{article}

 \usepackage{graphicx}

\begin{document}

\centerline{\large\bf Mixed scenario of the charged liquid surface
reconstruction}
 \centerline{V.Shikin and E.Klinovya}

 \centerline{\sl\small ISSP RAS, Chernogolovka, Russia}

\vskip 3 mm \
\begin{abstract}
Discussed in the paper is a mixed scenario of the charged liquid
surface reconstruction in the situation where the 2D surface
charge density is close to its saturation value.
\end{abstract}

One of the instabilities studied in classical hydrodynamics is the
Frenkel-Tonks (FT) instability [1,2] arising in a threshold manner
on the charged liquid surface and resulting in its deformation. In
contrast to other known instabilities, such as the Rayleigh
instability of a cylindrical jet [3], von Karman trace generated
by moving cylinder (sphere) [4], Taylor vortex instability of a
viscous layer between two co-axial rotating cylinders [5], etc.,
the FT decomposition process can be halted a new metastable state
with finite corrugation depth can be formed (charged liquid
surface reconstruction). The surface reconstruction was studied by
different authors at different times [6-19]. However, by the
present time a consistent comprehensive picture of the instability
development is only available for small values of the parameter
$\nu \ll 1$
$$
\nu=n_s/n_s^{max},\quad n_s^{max}=\sqrt{\alpha \rho g},
 \eqno(1)
$$
($\rho$ and $\alpha$ are the liquid density and surface tension,
$g$ is the gravity acceleration, $\kappa=\sqrt{\rho g/\alpha}$ is
the capillary length), determining the liquid boundary population
by localized charges. Explicitly manifested here is the dual
nature of the FT instability development: the spectral scenario
where the frequency imaginary part changes its sign as a function
of external parameters or, alternatively, or the surface
corrugation arises due to fluctuations. Under these conditions it
is possible to draw the phase diagram branches in the $(\nu, E_-)$
plane operating with the concepts used in the physics of the 1-st
order phase transitions: spinodal, binodal, etc. [20]. Here $E_-$
is the electric field above the charged liquid surface. At the
$\nu=1$ the electric field is $E_-=0$.

For finite values $\nu \le 1$ available in the theory are only
isolated facts which do not form any coherent picture. For
example, it is easy to extend the the spinodal in the $(\nu,
E_-)$-plane into the $\nu \le 1$ domain. Here the alternative part
of the problem, the nucleation-controlled ``decomposition'' of a
flat charged surface (the binodal) is left in the uncertain state
(there are no typical scale for the charge involved into formation
of a dimple). On the other hand, the experiment unequivocally
indicates a periodic reconstruction [18,19]. This difficulty is
lifted if one considers the FT instability temporal evolution.
Authors of Refs. [21,22] noted that the spinodal scenario of the
surface decomposition with a fixed total charge leads to the
appearance of a periodic array of neutral spots on charged
surface. This behavior obviously means the end of spino- and the
beginning of binodal evolution with well defined initial
conditions: the charge per arising dimple is known and all the
dimples are identical. We refer to the scenario beginning with the
``spino-'' regime and continuing with the ``bino-'' evolution of
the charged surface deformation as the mixed one. This scenario
reproduces in a qualitatively correct way basic known properties
of the FT instability in the $\nu \le 1$ domain: periodicity of
the reconstruction, loss of equipotentiality along the corrugated
liquid surface, and jump-like behavior of the corrugation
amplitude when the stationary reconstructed liquid boundary is
formed.

One should emphasize that the mixed decomposition scenario of
two-component systems is a rule rather than exception, the point
that has not yet been fully taken advantage of. Consider, for
example, the phase diagram in the vicinity of its extremum for a
two-component solution [23]. Standard system transition from a
``stable''point $a$ to unstable point $b$ by a jump of temperature
initiates the onset of the spinodal decomposition in the solution
along the $c-b-c_1$ isotherm. In this process the system
undergoing decomposition in the ``directions'' from $b$ to $c,c_1$
approaches the spinodal-binodal boundary from the inside of the
spinodal zone rather than from the uniform state which takes place
in the standard transitions of the $d-e$ type. Starting from the
$c^*-c_1^*$ points the evolution becomes binodal one characterized
by the spatially modulated stratified density arising at the end
of the spinodal decomposition phase (just like in the above
reconstruction problem) rather than by a random
nucleation-controlled density.

Turning back to the charged liquid surface, let us formulate a
number of arising problems. First of all, it is highly desirable
to make sure that the stationary ``dimpled'' state in
energetically more favorable than the unstable flat state
(experimentally, this is quite obvious since no deformation would
otherwise be observed). one should also estimate the charge
localization degree in the sense of the inequality
$$
\kappa R \le 1 ,\eqno(2)
$$
where $R$ is the single charged spot radius.

1.  In the standard flat capacitor geometry with the distance $h$
between the metal electrodes and the liquid film of thickness $d$
covered by a charged 2D system characterized by the number density
$n_s$ the problem of finding the surface properties is assumed to
be ``bulk-like'' if $h>d>>a$, where $a=\kappa^{-1}$ is the
so-called capillary length. In addition, we are only interested in
the saturation case where the electric field $E_-$ above the
liquid is zero while the density $n_s$ has its maximum possible
value, i.e. $\nu=1$. Under these conditions the electric field
$E_+^{max}$ in the liquid film also reached its maximum value
equal to
$$
E_+^{nax}=4\pi e n_s^{max} \eqno(3)
$$
where $\nu$ is taken from (1).

It is convenient to compare the system energy in different states
(flat $\bar W$ and corrugated $\tilde W$) by first considering the
collective effects and then calculating the contribution of a
single Wigner-Seitz cell. There are two collective effects. The
first one is the gain in Coulomb energy $\Delta w_c$ due to charge
ordering. This contribution plays the major role in Coulomb
crystallization on flat substrates and is estimated to be [24]
$$
\Delta w_c\simeq-1.1 Q^2/a \eqno(4)
$$
where $Q$ is the single dimple charge and $a$ is the lattice
spacing.

The second effect is specific to to the charged liquid surface. A
continuous electron disk of radius $R_*$ pressed by the electric
field $E_+^{nax}$ to the liquid surface squeezes some liquid into
the neutral neighborhood of its perimeter. This effect raises the
net electron system energy by $\Delta W_{\xi}$ (see Appendix). The
same disk driven to the lattice state displaces a different amount
of liquid (actually, it will be shown below that in that case no
liquid is displaced it all). Therefore, the surplus energy $\Delta
W_{\xi}$ favors the charged surface reconstruction.

In terms of $\Delta w_c$ and $\Delta W_{\xi}$ the general
requirement
$$
\bar W > \tilde W
$$
is written as
$$
\bar w_c +\Delta W_ {\xi}>\tilde w_c+n\tilde w_l, \quad  \bar
w_c-\tilde w_c=n\Delta w_c, \quad \Delta W_ {\xi}=n\Delta w_
{\xi}.\eqno(5)
$$
Here $\bar w_c$ is the total Coulomb energy of a capacitor with a
flat charged liquid layer, $\tilde w_c$ is the same energy in the
presence of a cluster lattice, $w_l$ is the total energy of a
single multiply charged dimple, $n>0$ is the total number of such
dimples arising in the reconstruction process. When writing Eq.
(5) we have assumed that in the mixed reconstruction the
corrugation possesses the following property: each cell of the
arising structure contains a single multiply charged dimple with a
charged core of radius $R$ such that $a/R\gg 1$ and the array of
dimples has a spatial period comparable to the capillary length.
This means that the inequality (5) we are interested in is
satisfied if
$$
\tilde w_l<\Delta w_c+\Delta w_{\xi}. \eqno(6)
$$

To estimate $w_l$ one could employ the available results [22] on
the properties of individual dimples. In that case the energy
$w_l$ becomes zero at
$$
E_{crit}=1.146 E_+^{max}
$$
which is qualitatively quite acceptable. However, the product
$\kappa R$ in that case yields $\kappa R\simeq 1.5
> 1$, making the single dimple approximation completely meaningless
in the calculations of $w_l$ for mixed reconstruction. Here the
problem should be reformulated from the very beginning taking into
account the finite size of the cell containing a single charged
spot.

The appropriate approach is given by the Wigner-Seitz (WS)
approximation with the external cell radius of the order of $a$
and the charged spot radius (actually, there exist two alternative
distributions differing by whether charged or empty spots occupy
the central part of the WS cell; qualitative estimates given in
the Appendix favor the charged central spots). In that case the
following set of equations is to be solved:
$$
eE_{\perp}\xi(r)+2\pi e^2\int\frac{n(r')r'dr'}{|r-r'|}= \lambda_e
 \eqno(7)
$$
$$
\Delta \xi -\kappa^2 \xi -P_{el}(r)/\alpha =\lambda_{\xi},\quad
P_{el}(r)=eE_{\perp}n(r) \eqno(8)
$$
$$
\partial\xi/\partial r|_{r\simeq a} =0, \quad 2\pi\int_0^a
rdr\xi(r)=0
 \eqno(9)
 $$
 $$
 2\pi\int n(r)rdr=N,\quad N=\pi a^2 n_s
 \eqno(10)
 $$
Here $\xi(r)$ is the self-consistent helium surface deformation,
$\kappa=\sqrt{\rho g/\alpha}$ is the liquid capillary length,
$P_{el}(r)=eE_{\perp}n(r)$ is the electron pressure on the helium
surface, $E_{\perp}$ is the electric field pressing charges
towards the liquid surface (in our case, according to Eq. (3),
$E_{\perp}=E_+=4\pi e n_s^{max}$), $\lambda_e$ is the Lagrange
multiplier accounting for the fixed total charge (10). Physically,
this multiplier is equivalent to $\tilde w_l$ (6). $\lambda_{\xi}$
is another Lagrange multiplier ensuring the total liquid volume
conservation (9).

Assuming the electron spot radius $R$ to be sufficiently small,
($\kappa R < 1$), it is convenient to represent the surface
deformation $\xi(r)$ in the central part of the cell as a series
$$
\xi(r)\simeq \xi(0)+\frac{1}{2}\xi^{\prime\prime}(0)r^2 + ...
\eqno(11)
$$
In that case, by analogy with the contact Hertz problem  [25]
$$
n(r)=\frac{3N}{2\pi R^2}(1-\frac{r^2}{R^2})^{1/2} \eqno(12)
$$
$$
E_{\perp}\xi^{\prime\prime}(0)=\frac{3}{2}Ne\int_0^{\infty}\frac{ds}
{(R^2+s)^2\sqrt{s}}\eqno(13)
$$
$$
\lambda_e-eE_{\perp}\xi(0)=\frac{3}{4}Ne^2\int_0^{\infty}\frac{ds}
{(R^2+s)\sqrt{s}}\eqno(14)
$$
In addition, in the central part of the cell the Eq. (8) becomes
$$
\xi^{\prime\prime}(0)-\kappa^2 \xi(0)-\lambda_{\xi}=
3QE_{\perp}/(4\pi R^2\alpha), \eqno(15)
 $$

Relations (13) and (15) contain there dimple characteristics:
$\xi^{\prime\prime}(0),~R,~\xi(0)$. To close this set of
equations, one should add a definition of $\xi(0)$, i.e. the
solution of (8) which can be obtained by expanding $\xi(r)$ into a
series in appropriate Bessel functions. By writing
$$
\xi(r)=\bar\xi+\tilde\xi(r),\quad \bar\xi=-\lambda_{\xi},\eqno(16)
$$
and choosing the functions
$$
J_0(\mu_nr/a), \quad \frac{\partial J_0(\mu_nr/a)}{\partial
r}|_{r=a}=0
$$
as a basis set so that to automatically satisfy the first boundary
condition in (9) one can obtain to solution to Eq. (8) as a series
$$
\xi(r)=\bar\xi+\sum_{n=1}^{\infty}\tilde\xi_n
J_0(\mu_nr/a),\eqno(17)
$$
$$
\tilde\xi_n=-\frac{A f(\mu_n,R/a)}{1+\mu_n^2J_0^2(\mu_n)},\quad
f(x)=\frac{sin x-xcos x}{x^3}, \quad
A=\frac{3QE_{\perp}}{\pi\alpha}.
$$
The quantity $\bar\xi$ here is obtained from the second boundary
condition (9) and proves to be zero.

Analysis of the equation set (13), (15), (17) allows one to
suggest (see arguments cited in the Appendix) that at the initial
stage of the dimple reconstruction development the quantity
$\xi(0)$ does not play any significant role in these equations.
Hence, the relationships between $\xi^{\prime\prime}(0)$ and $R$
immediately yield a closed equation set from which they can be
easily found:
$$
E_{\perp}\xi^{\prime\prime}(0)=\frac{3\pi Q}{4 R^3}, \quad
R=\pi^2\alpha/E_{\perp}^2,\eqno(18)
$$
Here the quantity $\xi(0)$ is obtained from Eq. (17) with $R$
calculated from Eq. (18).

The final stage in the estimation of the mixed scenario
feasibility is the estimation of parameters which are most
critical for its realization. By setting in the general formulas
(7)-(18) $E_{\perp}=E_+=4\pi e n_s^{max}$,  one has
$$
\frac{R_{crit}}{a}\simeq 0.4<1,\quad \frac{\xi_{crit}(0)}{a}\simeq
0.3<1,\quad\lambda_e^{crit}\equiv \tilde w_l^{crit}< \Delta
w_c^{crit}+\Delta w_{\xi}^{crit} \eqno(19)
$$
Here $n_s^{max}$ is taken from (1) and $\tilde w_l$ from (6).

Estimates (19) reveal that the mixed scenario of the charged
liquid helium surface reconstruction seems to be rather acceptable
in the vicinity of the filling factors (1) close to unity. In the
present paper, it is impossible to obtain more definite
conclusions since the employed approximation (B-3) cannot claim
any quantitatively correct statements. We do not see any other
approaches (different from the mixed scenario) which could proved
at least qualitative explanation of the observed periodic
reconstruction of the liquid helium surface in the range of $\nu
\le 1$.

This work was partly supported by the RFBR.

\centerline{\sl\small Appendix}

 1. Let $R_*$ be the charged area radius and $L>R_*$ the radius of entire liquid surface
between vertical walls. The equilibrium of the cell as a whole is
defined by the following equations:
$$
\rho g\xi_{0}+P_{el}=\rho g \xi_{1}, \eqno(P1)
 $$
$$
R_*^2\xi_{0}+(L^2-R_*^2)\xi_{1}=0, \quad L > R_*.\eqno(P2)
$$
Here $ \xi_{0}$ is the liquid surface deformation in the charged
area, $a, \xi_{1}$ is the the liquid surface deformation outside
the electron disk, and $P_{el}$ is the effective pressure in the
charged area.

Finding from (P2) the quantity $\xi_{1}$
$$
\xi_{1}=-\xi_{0}R_*^2/(L^2-R_*^2)
$$
and substituting it into (P1) one has
$$
\xi_{0}=-P_{el}/g^*, \quad g^*=g(1+\frac{R_*^2}{L^2-R_*^2} )
\eqno(P3)
$$

The additional energy $\Delta W_{\xi}$ due to deformation
$\xi_{0}$ has the scale
$$
\Delta W_{\xi}\simeq \pi R_*^2 P_{el}\xi_{0} \eqno(P4)
$$
Combining all the necessary definitions one obtains from (P4)
$$
\Delta W_ {\xi}=n\Delta w_ {\xi}, \quad\Delta w_
{\xi}=\frac{4\pi\alpha a^2}{f(R_*,L)},\quad
f(R_*,L)=(1+\frac{R_*^2}{L^2-R_*^2} ).\eqno(P4)
$$
This expression is used for estimates in the main text of the
paper.

2. Within the alternative including either charged or empty spots,
it is convenient to start from qualitative estimates. Suppose we
consider a cylindrical Wigner-Seitz model with the spot external
radius of the order of $a$, charged spot radius $R<a$, and
charge-free ring with area $\pi(a^2-R^2)$. If the last formula is
interpreted as an integral relation, one can estimate within the
cylindrical model the probability of scenario described in Ref.
[21] where the initial stage of the reconstruction is assumed to
be represented by a periodic array of empty (charge-free) spots
whose area is different from zero to the extent of positive
supercriticality $\delta E= (E_- -E_-^c)$,
$$
\delta E= (E_--E_-^c). \eqno(P5)
$$
However, this is most probably not so because of the following
reasons.

Within the WS cell both the local equilibrium conditions and
liquid volume conservation should be satisfied, just as in (P1).
Then, by similar arguments one has
$$
\xi_{0}=-P_{el}/g^*, \quad g^*=g(1+\frac{R^2}{a^2-R^2} ) \eqno(P6)
$$
According to (P6), development of a stationary corrugation under
the conditions $\pi(a^2-R^2)\to 0$ (empty spots with low area) is
unlikely since in that domain the effective value of
$g^*\to\infty$. In that case the appearance of corrugation, i.e.
growth of $\xi_{0}$ (P6) is hindered.


\begin{thebibliography}{00}

\bibitem{1} Ya.Frenkel,  Zs. der Sowietunion {\bf 8}, 675,
(1935); Ya.Frenkel, ZhETF  {\bf 6},  347, (1936).

\bibitem{2} T.Tonks. Phys. Rev. {\bf 48}, 562 , (1935).

\bibitem{3} in G.J.W.S.Rayleigh and R.B.Lindsay, \textit{The Theory of Sound, Volume 2.} Dover
Publications, 1976.

\bibitem{4} G.K.Batchelor. \textit{An Introduction to Fluid Dynamics.} Cambridge University Press, 1967.

\bibitem{5} L.D.Landau and E.M.Lifshitz. \textit{Fluid Mechanics}. Pergamon Press, 2nd ed., 1987.

\bibitem{6} J. Melcher. \textit{Field-coupled Surface Waves}, Cambridge, Mass. The
MIT Press 1963.

\bibitem{7} G.Taylor, A.McEwan. J.Fluid Mech. {\bf 22}, 1, (1965).

\bibitem{8} M.Cowley, R.Rosenswieg. J.Fluid Mech. {\bf 30}, 671, (1967).

\bibitem{9} V.Zaitsev, M.Shliomis.  Doklady USSR Acad . Sci., {\bf 188}, 1261, (1969).

\bibitem{10} L.Gor'kov, D.Chernikova.  Pis'ma  ZhETF {\bf 18}, 119, (1973).

\bibitem{11} M.Shliomis. Usp. Phys. Nauk,  {\bf 112}, 437, (1974).

\bibitem{12} D.Chernikova. ZhETF {\bf 68}, 250 (1975).

\bibitem{13} E.Kuztetsov, M.Spektor. ZhETF {\bf 71}, 262, (1976).

\bibitem{14} A.Volodin, M.Khaikin, V.Edel'man. Pis'ma ZhETF {\bf 23}, 524,
(1976).

\bibitem{15} L.Gor'kov, D.Chernikova.  Doklady USSR Acad . Sci., {\bf 228},  829, (1976).

\bibitem{16} A.Volodin, MKhaikin, V.Edel'man. Pis'ma ZhETF {\bf 26}, 707,
(1977).

\bibitem{17} P.Leiderer. Phys. Rev. {\bf B 20}, 4511, (1979).

\bibitem{18} P.Leiderer, M.Wanner. Phys.Lett.  {\bf A 73},  189, (1979).

\bibitem{19} M.Wanner, P.Leiderer. Phys. Rev. Lett. {\bf 42}, 315, (1979).

\bibitem{20} V.Shikin. Pisma ZhETF {\bf 78} 930 (2003).

\bibitem{21} V.I.Mel'nukov, S.V.Meshkov.  Pis'ma ZhETF {\bf 33}, 222,
(1981).

\bibitem{22} V.I.Mel'nukov, S.V.Meshkov.  ZhETF  {\bf 81}, 951, (1981).

\bibitem{23} A.Olemskii, I.Koplyk. Usp. Phys. Nauk, (1995), 1105

\bibitem{24} L.Bonsall, A.Maradudin.  Phys. Rev. {\bf B15}, (1977), 1959

\bibitem{25} L.D.Landau and E.M.Lifshitz. \textit{Theory of
Elasticity.} Pergamon Press, 3rd ed., 1986.

\end{thebibliography}
\end{document}